# The Promise of Data Science for the Technosignatures Field

**Thematic Areas:**   v Methods

**Principal Author:**
Name: Anamaria Berea
Institution: University of Central Florida
Email: anamaria.berea@ucf.edu
Phone: 571-314-3768

**Co-authors:**
Name: Steve Croft
Institution: University of California, Berkeley
Email:  scroft@astro.berkeley.edu
Phone:  925-321-0871

Name: Daniel Angerhausen
Institution: Center for Space and Habitability
Email: daniel.angerhausen@gmail.com
Phone: +41 77 463 53 41

**Abstract**

This paper outlines some of the possible advancements for the technosignatures searches using the new methods currently rapidly developing in computer science, such as machine learning and deep learning. It also showcases a couple of case studies of large research programs where such methods have been already successfully implemented with notable results. We consider that the availability of data from all sky, all the time observations paired with the latest developments in computational capabilities and algorithms currently used in artificial intelligence, including automation, will spur an unprecedented development of the technosignatures search efforts.

**Introduction**

Data science is a rapidly evolving field that includes advanced computational methods and capabilities, such as machine learning and deep learning algorithms, as well as increased computational power and data storage capabilities on servers and data centers. Some of the striking results in various scientific fields derived from the application of deep learning algorithms have shown improved predictability and forecasting accuracy over previous methods, as well as pattern recognition, classification and reclassification of events or images, ontological discoveries, and many more. In the technosignatures field, these new methods can bring insights in two ways: 1. Through a re-analysis and re-examination of existing data from both radio and optical astronomy using new algorithms; 2. Through data simulations or data proxies from similar fields (i.e., cryptology, cybersecurity, datasets with low signal to noise ratio) where these data would be manipulated and siphoned through advanced computational techniques as well. The purpose of both these approaches would be to discover new patterns if such patterns have been overlooked, to derive quick inferences about outliers in the data, to categorize and classify the data in the least biased way possible (unsupervised), as well as to improve the time frame and the amount of data to be analyzed in an automatic/semi-automatic fashion for signal to noise detection. And not last, the promise of AI derived algorithms would be to ultimately converge to a standardized tool/methodology that would be adaptive to different types of data (radio, image, etc.) and render probabilistically the type of signal detected. While the training of any machine learning or classifying system requires large datasets and a great deal of computational effort, once established these detection criteria are easy to apply, and may be the only way that we can approach the requisite real-time analysis as observational data rates continue to increase and different observing opportunities proliferate, approaching all-sky, all-the-time.

This paper presents two major efforts done at the institutional level with respect to the use, research and development of machine learning and deep learning algorithms with the specific purpose of searching for biosignatures and technosignatures. One is represented by NASA Frontier Development Lab, a research accelerator and the other one by the Breakthrough Foundation.

**NASA/SETI Frontier Development Lab (FDL)**

One example for a potential future unique industry/academia collaboration is the NASA Frontier Development Lab (https://frontierdevelopmentlab.org). In the past three years, the 8 week long program developed in partnership with NASA's Ames Research Center at the SETI institute tackled problems ranging from predicting solar storms to finding life in space. The program aims to apply AI technologies to challenges in space exploration by pairing machine learning expertise with space science and exploration researchers from academia and industry. These interdisciplinary teams address tightly defined problems and the format encourages rapid iteration and prototyping to create outputs with meaningful application to the space program and humanity. What makes FDL unique is its close collaboration with industry leaders in AI, such as Intel, Google, Kx Systems, IBM and NVIDIA and key players in private space such as SpaceResourcesLu, Lockheed Martin, KBRWyle and XPRIZE. In 2018 FDL focused on two challenges from the wide field of Astrobiology and one challenge about Exoplanets.

The Exoplanet team used state-of-the-art deep learning models to automatically classify Kepler transit signals as either exoplanets or false positives (Ansdell et al., 2018). Their Astronet code expanded upon work of Shallue and Vanderburg (SV18) by including additional scientific domain knowledge into the network architecture and input representations to significantly increase overall model performance. Notably, they achieved 15 - 20 percent gains in recall for the lowest signal-to-noise transits that can correspond to rocky planets in the habitable zone. They input CCD pixel centroid time-series information derived from Kepler data and key stellar parameters taken from the catalogue into the network and also implement data augmentation techniques to alleviate model over-fitting. These improvements allowed them to drastically reduce the size of the model, while still maintaining improved performance. These smaller models are better for generalization, for example from Kepler to TESS data. Their work illustrates the importance of including expert domain knowledge in even state-of-the-art deep learning models when applying them to scientific research problems that seek to identify weak signals in noisy data. This classification tool will be especially useful for upcoming space-based photometry missions focused on finding small planets, such as TESS and PLATO (e.g., Osborn et al. 2019).

In Soboczenski et al. 2018 and Cobb/Himes at al. 2019, the Astrobiology 2 team presented a Machine-Learning-based retrieval framework called Intelligent exoplaNet Atmospheric RetrievAl (INARA) that consists of a Bayesian deep learning model for retrieval and a data set of 3,000,000 synthetic rocky exoplanetary spectra generated using the NASA Goddard Planetary Spectrum Generator (PSG). This work represents the first ML retrieval model for rocky, terrestrial exoplanets and the first synthetic data set of terrestrial spectra generated at this scale (Zorzan et al, in prep; O'Beirne et al., in prep).

The Astrobiology team 1 project demonstrated how cloud computing capabilities can accelerate existing technologies and map out previously neglected parameter spaces (Bell et al., 2018). They succeeded in modelling tens-of-thousands of potential atmospheres over a few days, using software ATMOS that was originally intended for use in single run applications. The full atmospheric composition data set that was generated will become a useful resource for the community to understand distributions of habitability parameters such as surface temperatures, photon and redox free energy availability for different classes of planetary systems and will enable better interpretations of future observations of exoplanet atmospheres and potential biosignatures.

The software product created during FDL has the potential to significantly improve the accessibility of ATMOS for a wide community of researchers, particularly for researchers with limited experience in handling FORTRAN source code. The parameter search approach, enabled by GCP, can be adopted by the community to simulate more atmospheres and/or modified to rapidly iterate on other problems that utilize ATMOS. More generally, the project showed that the approach of containerizing legacy or hard-to-use software, executing it over massively parallel processing, and setting up a procedural execution algorithms using the Google Cloud Platform could become a template for similar parameter search problems in and beyond astrobiology.

**Breakthrough Listen**

One of the largest searches for technosignatures currently underway is the *Breakthrough Listen* initiative (Worden et al. 2017). *Listen* is using around 20% of the time on the Green Bank Telescope (GBT) and Parkes Telescope, in addition to an optical search using the Automated Planet Finder, and partnerships in various stages of development with other facilities around the globe. Much of the focus of the GBT program to date has been targeted searches of nearby stars for Doppler-drifting artificial signals (Enriquez et al. 2017), but the program has also delivered new insights into astrophysical phenomena including fast radio bursts (FRBs).

In addition to using classical algorithms to search for both technosignature candidates and FRBs, the *Listen* team has been developing machine learning approaches to find signals that otherwise would have been missed, to aid in classification, and to search for anomalies. Zhang et al. (2018) trained a convolutional neural network (CNN) on simulated FRBs, and then ran the network on 400 TB of raw GBT data from the repeating FRB 121102. Gajjar et al. (2018) had previously reported 21 bursts in this dataset found using a classical tree dedispersion algorithm; Zhang et al.'s CNN found 72 bursts that had previously not been detected.

Zhang et al. (2019) developed a self-supervised anomaly detection method for narrowband searches, using an adversarial convolutional long short-term memory (LSTM) network to predict the time evolution of detected signals. The algorithm is capable of detecting anomalies in observations of target stars that are not seen in comparison observations (Fig. 1).

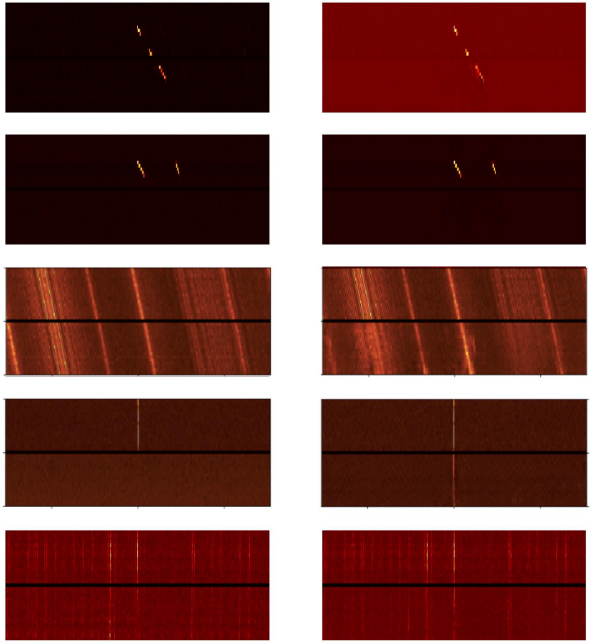

Fig. 1. Example reconstruction and predictions (right) of GBT waterfall plots (spectrograms) of 5 pairs of "ON-OFF" observations from the network developed by Zhang et al. (2019) compared to actual observations (left). The network does a good job of correctly predicting the "OFF" observation (bottom half of each image) from the "ON" observation, except in panel 4 of the left column, where a signal seen in the "ON" data is not present in the "OFF". This anomaly is consistent with how a technosignature would be expected to appear in the data.

**BIg Data on Earth and in the Universe**

In 2015, it was estimated that approximately 4.7 exabytes of data have been generated by the astronomical observatories of various sky survey projects (Zhang et al., 2015). By comparison, humans generate approximately 2.5 exabytes of data every day (Marr, 2018). Most of the current algorithms and research regarding Big Data and machine learning techniques have been developed using the noisy and high volumes of the data we are currently generating daily and that is easily accessible. When translated to the cleaner, instrument collected data, these algorithms and methods tend to perform even better.

One research where the performance of these algorithms is shown comparatively is in Harper et al., 2019. The authors used time series voltage data, transformed it into 2D representations such as waterfall plots and used convolutional neural networks usually developed for image recognition in order to extract information about the different types of signals (Harper et al., 2019). Additionally, the authors used principal component analysis extraction too. This method showed good performance, similar to the traditional detection algorithms, but with the advantage of being automatable and usable for very large sets of data.

**Conclusions and Recommendations**

Data science is a relatively new field, that is mostly morphing from the rapid developments in data collection capabilities and algorithmic and computational innovations. Data science today is mostly associated with the use of very large datasets (continuously being collected) siphoned through advanced machine learning and deep learning algorithms with the purpose to either improve predictability or to discover patterns of the phenomenon under scrutiny.

Some of the conclusions we can immediately draw:
- Data science is a rapidly evolving field with impressive results in fields such as physics, biology, social sciences, and many more
- Current research that yielded impressive results from the application of machine learning algorithms are NASA Frontier Development Lab (in biosignatures search) and Breakthrough Listen (in technosignatures search).
- All sky, all the time will make available even more data that is suitable for automation and the newest algorithms currently used in AI; in fact, it will be impossible to analyze all these data accurately and in a timely manner.